\documentclass[prd,twocolumn]{revtex4}
\usepackage{graphicx, epsfig}
\usepackage{color}
\usepackage{mathrsfs}

\newcommand{\be}{\begin{equation}}
\newcommand{\ee}{\end{equation}}
\newcommand{\bea}{\begin{eqnarray}}
\newcommand{\eea}{\end{eqnarray}}

\newcommand{\gapp}{\mathrel{\raise.3ex\hbox{$>$}\mkern-14mu
              \lower0.6ex\hbox{$\sim$}}}
\newcommand{\lapp}{\mathrel{\raise.3ex\hbox{$<$}\mkern-14mu
              \lower0.6ex\hbox{$\sim$}}}

\begin{document}
\title{Schrodinger Picture of Quantum Gravitational Collapse}
\author{Tanmay Vachaspati}
\affiliation{Institute for Advanced Study, Princeton, NJ 08540.\\
CERCA, Department of Physics, 
Case Western Reserve University, Cleveland, OH~~44106-7079.}

\begin{abstract}
\noindent
The functional Schrodinger equation is used to study the quantum 
collapse of a gravitating, spherical domain wall and a massless 
scalar field coupled to the metric. The approach includes 
backreaction of pre-Hawking radiation on the gravitational 
collapse. Truncating the degrees of freedom to a minisuperspace 
leads to an integro-differential Schrodinger equation. We define 
a ``black hole'' operator and find its eigenstates. The black hole 
operator does not commute with the Hamiltonian, leading to an 
energy-black holeness uncertainty relation. We discuss energy
eigenstates and also obtain a partial differential equation for 
the time-dependent gravitational collapse problem.
\end{abstract}

\maketitle

\section{Introduction}
\label{introduction}

Gravitational collapse is expected to result in the
formation of a black hole, which then evaporates by Hawking
radiation \cite{Hawking:1974sw}. During this process, any 
information contained in the initial state is lost, since 
black hole evaporation leads to thermal 
radiation which is uncorrelated with the initial state. 
If the initial state is chosen to be a pure quantum state, 
the final state would be described by a density matrix 
and would not be a pure state. The evolution of a pure quantum
state into a mixed state violates the tenets of quantum
mechanics ({\it e.g.} \cite{Susskindbook}). These problems of 
black hole formation, evaporation and information loss have been 
central to discussions of combining general relativity and 
quantum mechanics over the last several decades. 

A number of approaches have been made to resolve the issues
that arise in combining black holes and quantum field theory,
including more sophisticated calculations of the emitted
radiation, modifications of quantum mechanics, modifications 
of quantum field theory, lower dimensional calculations, loop 
gravity, and string theory ({\it e.g.} for some reviews see 
\cite{Susskindbook,frolov,Strominger:1994tn,Wald:1998de,
Ashtekar:2004eh}). 
Depending on the approach and the
new ingredient in the calculation, the conclusions have varied, 
and no concensus has emerged so far. The issues have become ever 
more pressing with discussion of black hole formation in highly 
energetic collisions in 
LHC \cite{Banks:1999gd,Giddings:2001bu,Dimopoulos:2001hw}
and by ultra-high energy cosmic rays colliding with the 
atmosphere \cite{Anchordoqui:2001cg}.

In contrast to earlier studies in which a black hole spacetime
is adopted as an arena for quantum field theory, we wish to
study the temporal development of an initial state that 
does not contain a black hole. From the laws of quantum
mechanics, the initial state will evolve unitarily in time.
If unitary evolution leads to black hole formation then,
as has been described by Hawking and others 
\cite{Hawking:1974sw,Susskindbook}, unitarity will be lost.

\begin{figure}
\scalebox{0.6}{\includegraphics{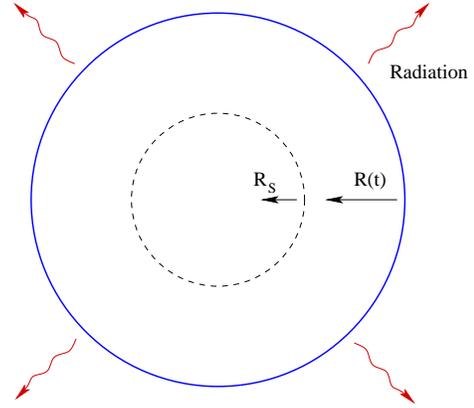}}
\caption{A picture of the system under consideration. A spherical
domain wall with radius $R(t)$ collapses towards its Schwarzschild 
radius, $R_S$, and emits scalar radiation in the process. Due to
the emission, $R_S$ also decreases.
}
\label{systemfig}
\end{figure}

The 3+1 dimensional analysis in Ref.~\cite{Vachaspati:2006ki} 
used the functional Schrodinger formalism to study the temporal
evolution of a scalar field in the {\it classical} background 
of a gravitating, collapsing, spherical domain wall. It was found
that the collapse was accompanied by quantum radiation of scalar 
particles, called ``pre-Hawking'' radiation, with a non-thermal
spectrum. 
The analyses in Refs.~\cite{Vachaspati:2006ki,Vachaspati:2007hr} 
were clearly limited by their use of the semiclassical approximation, 
in which the backreaction of the scalar radiation on the 
gravitational collapse spacetime is ignored. The purpose of 
the present work is to attempt to remedy 
this shortcoming within the functional Schrodinger formalism.
The aim of the present work is to provide a framework to study
quantum gravitational collapse beyond the semiclassical approximation.
The framework includes backreaction of the scalar radiation on the
collapse dynamics (see Fig.~\ref{systemfig}).

The specifics of the domain wall plus scalar field system are not 
expected to be very important for the questions we are interested 
in. For example, the Hamiltonian for a collapsing wall that is 
close to forming a black hole reduces to the form 
for an ultra-relativistic particle \cite{Vachaspati:2006ki}, which
is expected to be true for any form of matter. So the choice of
a collapsing domain wall does not affect the near-horizon dynamics.
The scalar field can also be thought of as representing one degree 
of freedom of a photon or a graviton. One assumed property that is 
special though, is the assumed masslessness of the scalar field. 
If the scalar field has a mass and/or carries a global or gauge 
charge, we do expect some differences that we discuss briefly in 
Sec.~\ref{discussion}. 

In general, the Schrodinger formulation yields a functional
differential equation for the wavefunctional,
$\Psi [g_{\mu\nu}, X^\mu , \Phi, t]$, where $g_{\mu\nu}$ is the
metric, $X^\mu$ the domain wall position, and $\Phi$ the
scalar field. We will truncate this problem to minisuperspace,
considering only spherical domain walls described by a
radial function, $R(t)$, the metric fixed to the
Schwarzschild form though with variable Schwarzschild
radius, $R_S$, and $\Phi$ decomposed into modes with
mode coefficients $\{ a_k \}$. Then the wavefunctional gets
replaced by an ordinary wavefunction, $\Psi (R,\{ a_k \},t)$.
We are interested in finding the time evolution of this
minisuperspace wavefunction, starting from black hole free
initial conditions, and in determining if there is a breakdown 
of the evolution at some time.

In any treatment of the gravitational collapse problem, it is 
necessary to choose a time coordinate since the collapse, by 
its very definition, is an evolution problem. In the current
analysis, we have chosen to work with the Schwarzschild time 
coordinate. 
However, there is the possibility that we may miss some portion
of spacetime due to this choice of slicing. If this were the 
case, we would expect some sickness ({\it e.g.} geodesic 
incompleteness) to show up in the temporal evolution of the 
wavefunction. If the temporal evolution remains well-behaved 
at all times, the evolution is no different from that of an 
ordinary star for which Schwarzschild coordinates are valid 
and commonly used. In other words, the Schrodinger formulation 
simply evolves the system forward in time, without assuming 
the presence or absence of an event horizon in the future. 
This is different from other approaches where quantum field 
theory is used on a spacetime background, where the entire 
spacetime must be assumed at the onset and then sliced using 
some coordinate system. However, it would be worthwhile to 
re-work gravitational collapse in the Schrodinger formalism 
with a different choice of time coordinate. 

The backreaction problem, even for classical point charges, is 
notorious for its non-local nature, since the trajectory depends 
on the radiative losses over the entire past. However, this problem
only occurs in a perturbative treatment of a point charge, since
there is then a zeroth order trajectory due to which there
is radiation, and then backreaction to first order in some
coupling constant, then a first order correction to the
trajectory, then radiation and backreaction to second order,
{\it ad infinitum}. Instead, if the classical point charge
is replaced by a regular solution to some field theory, as
in a 't Hooft-Polyakov magnetic monopole, a classical solution
of the field theory will include the full dynamics and radiation
of the point (magnetic) charge. Similarly, the wavefunction 
for the spherical wall and radiation in the Schrodinger 
formalism, includes the full dynamics of the wall and the
radiation and non-locality is absent for this reason. However, 
there are two other reasons that make the analysis difficult
and lead to non-locality. The first is that the pre-Hawking
scalar radiation not only affects the dynamics of the collapse
but also contributes to the precise form of the metric. In
the minisuperspace approach we have adopted, a form of the
metric needs to be specified. This is chosen to be of the
Schwarzschild form and thus amounts to the assumption that
the energy-momentum tensor in the pre-Hawking radiation
only causes negligible departures from the minisuperspace.
We expect this approximation to be justified for large
collapsing mass where we know that the energy-momentum
density of the radiation is comparitively small. This has also 
been explicitly verified in the semiclassical calculation
\cite{Davies:1976ei}. Also, the present approach is similar 
to that taken in calculating radiation backreaction on an
accelerating charge in electrodynamics where the backreaction
is taken to affect the dynamics of the charge but the effects
on the Coulomb electric field of the charge is not considered.
The second factor that complicates 
the analysis is the non-linear nature of gravity. In 
particular, the minisuperspace Hamiltonian depends on the
mass of the collapsing object, and this is itself related to
the Hamiltonian. In order to isolate the Hamiltonian, we
need to invert the momentum operator and this leads to an
integro-differential form for the Hamiltonian. Solutions of
the corresponding integro-differential Schrodinger equation
are hard to find in general but we are able to transform 
the problem to a purely differential equation in the 
``incipient limit'' where the collapse approaches black hole 
formation.  We discuss energy eigenstates in 
Sec.~\ref{stationary} and the time-dependent gravitational
collapse problem in Sec.~\ref{gravcollapse}. 

In order to solve the gravitational collapse problem, we need
to solve the Schrodinger equation to obtain a wavefunction that 
describes the collapsing wall and radiation. The first task, 
however, is to specify an initial value for the wavefunction,
such that the initial state itself does not contain a black hole. 
This means that we need to specify a criterion for deciding if 
a given wavefunction is black hole free. To answer this question, 
we propose a ``black hole operator'' in Sec.~\ref{bhfstate}. 
Eigenvalues of the black hole operator signify the ``black
holeness'' of the state. Interestingly we find that the
black hole operator does not commute with the Hamiltonian,
implying an uncertainty relation between energy and black 
holeness. 

We start by discussing the classical Hamiltonian 
(Sec.~\ref{classicalhamiltonian}), which is then promoted to
a quantum Hamiltonian in Sec.~\ref{quantumhamiltonian}. 
We then discuss explicit representations of an important
operator in Sec.~\ref{hatB} and use it to define a
black hole free state in Sec.~\ref{bhfstate}. We then discuss
energy eigenstates in Sec.~\ref{stationary} and the gravitational
collapse problem in Sec.~\ref{gravcollapse}.
We discuss our results and conclude in Sec.~\ref{discussion}.
A discussion of Hermiticity of certain operators can be
found in Appendix~\ref{hermiticity}.

\section{Classical Hamiltonian}
\label{classicalhamiltonian}

The action contains the Einstein-Hilbert, massless scalar field,
and Nambu-Goto terms
\begin{equation}
S = \int d^4 x \sqrt{-g} \left [ -\frac{\cal R}{16\pi G}
     + \frac{1}{2} (\partial_\mu \Phi )^2 \right ]
     - {\sigma} \int d^3 \xi \sqrt{- \gamma}
\label{action}
\end{equation}
where $\sigma$ is the wall tension and the domain wall world-volume
metric is given by
\begin{equation}
\gamma_{ab} = g_{\mu\nu} \partial_a X^\mu \partial_b X^\nu
\label{inducedmetric}
\end{equation}
We will restrict our attention to spherical symmetry, in which
case the form of the line element for the domain wall alone is
\cite{Ipser:1983db}
\begin{equation}
ds^2= -(1-\frac{R_S}{r}) dt^2 + (1-\frac{R_S}{r})^{-1} dr^2 +
      r^2 d\Omega^2 \ , \ \ r > R(t)
\label{metricexterior}
\end{equation}
where, $R_S = 2GM$ is the Schwarzschild radius in terms of the mass,
$M$, of the wall, and $d\Omega^2$ is the usual angular line element.
In the interior of the spherical domain wall, the line element
is flat, as expected by Birkhoff's theorem,
\begin{equation}
ds^2= -dT^2 +  dr^2 + r^2 d\theta^2  + r^2 \sin^2\theta d\phi^2  \ ,
\ \ r < R(t)
\label{metricinterior}
\end{equation}
\begin{equation}
\frac{dT}{d\tau} =
      \left [ 1 + \left (\frac{dR}{d\tau} \right )^2 \right ]^{1/2}  ,
 \ \ \
\frac{dt}{d\tau} = \frac{1}{B} \left [ B +
         \left ( \frac{dR}{d\tau} \right )^2 \right ]^{1/2}
\label{bigTandtau}
\end{equation}
\begin{equation}
B \equiv 1 - \frac{R_S}{R}
\label{BofR}
\end{equation}

We will consider
the case when the mass of the domain wall is large compared to 
the energy-momentum contribution of the scalar field and so 
it is a good approximation to ignore the scalar field when
writing the metric. 

Our next goal is to find the Hamiltonian for the wall-scalar field
system. In Ref.~\cite{Vachaspati:2006ki} we have found that the mass of the
wall can be written as
\begin{equation}
M = 4\pi \sigma R^2 \left [ \frac{1}{\sqrt{1-R_T^2}} 
                           - 2\pi G\sigma R \right ]
\end{equation}
or more suggestively
\begin{equation}
M = \frac{\tilde M}{\sqrt{1-R_T^2}} - \frac{G {\tilde M}^2}{2R}
\end{equation}
where ${\tilde M} \equiv \sigma 4\pi R^2$. The first term is
the Lorentz boosted energy contribution while the second is
the gravitational binding energy.
Using this expression for the mass and the relation between
$T$ and $t$ in Eq.~(\ref{bigTandtau}) we find
\begin{equation}
H_{\rm wall} = 4\pi \sigma B^{3/2}R^2 \left [
         \frac{1}{\sqrt{B^2-{\dot R}^2}} -
          \frac{2\pi G\sigma R}{\sqrt{B^2- (1-B){\dot R}^2}}
                                  \right ]
\label{energyfunction}
\end{equation}
where overdots denote derivatives with respect to $t$.

The form of the wall Hamiltonian simplifies in the incipient limit
($B \to 0$). Then the canonical momentum is given by 
\begin{equation}
\Pi \approx \frac{4\pi \mu R^2  {\dot R}}
              {\sqrt{B} \sqrt{B^2-{\dot R}^2}}
\end{equation}
where $\mu \equiv \sigma (1-2\pi G\sigma R_S)$, leading to
\begin{eqnarray}
H_{\rm wall}
&\approx& \frac{4\pi \mu B^{3/2}R^2}{\sqrt{B^2-{\dot R}^2}} \label{HRdot}\\
  &=& \left [  (B\Pi)^2 + B (4\pi \mu R^2)^2 \right ] ^{1/2} \label{HPi}
\end{eqnarray}
which has the form of the energy of a relativistic
particle, $\sqrt{p^2 + m^2}$, with a position dependent mass.
In the limit $B \to 0$, the mass term can be neglected -- the
wall is ultra-relativistic -- and hence
\begin{equation}
H_{\rm wall} \approx - B \Pi
\label{finalHwall}
\end{equation}
where we have chosen the negative sign appropriate for describing
a collapsing wall.

Next we introduce the scalar field $\Phi$. Even when we include 
the scalar field, we will continue to use the metric in the 
Ipser-Sikivie form described above. This assumes that the dominant 
effect of backreaction on the metric is to change the wall energy
which enters the metric via $R_S$. This is not rigorously true since 
the scalar field also contributes a non-vanishing energy-momentum 
density. However, we assume that this contribution is small compared 
to the energy in the wall. In the conventional case of evaporation
from an existing black hole, this corresponds to the assumption
that Hawking radiation causes the black hole to evaporate and
lose mass, but the black hole metric remains Schwarzschild to
a very good approximation.

The scalar field, $\Phi$, is decomposed into a complete set
of basis functions denoted by $\{ f_k (r)\}$
\begin{equation}
\Phi = \sum_k  a_k(t) f_k (r)
\label{modes}
\end{equation}
The exact form of the functions $f_k (r)$ will not be important for us.
We will be interested in the wavefunction for the mode coefficients
$\{ a_k \}$.

The Hamiltonian for the scalar field modes is found by
inserting the scalar field mode decomposition and the background
metric into the action
\begin{equation}
S_\Phi = \int d^4x \sqrt{-g} \frac{1}{2} g^{\mu \nu}
                 \partial_\mu \Phi \partial_\nu \Phi
\end{equation}
The Hamiltonian for the scalar field modes takes the form of coupled
simple harmonic oscillators with $R-$dependent masses and couplings due
to the non-trivial metric. In the regime $R \sim R_S$, for a normal
mode denoted by $b$, the Hamiltonian is \cite{Vachaspati:2006ki}
\begin{equation}
H_{\rm b} = \left ( 1- \frac{R_S}{R} \right ) \frac{\pi^2}{2m}
           + \frac{K}{2} b^2
\end{equation}
where $\pi$ is the momentum conjugate to $b$, and $m$ and $K$ are
approximate constants whose precise values are not important for 
us. The reason $m$ and $K$ are only approximately constant is 
because they depend on $R_S$ and, with backreaction included,
$R_S$ changes (slowly) with time. By treating $m$ and $K$ as
constant, we are assuming that the dominant coupling between
$b$ and $R$ is due to the $(1-R_S/R)$ factor.

Hence the total Hamiltonian for the wall and the normal
modes of the scalar field in the incipient limit is
\begin{equation}
H = H_{\rm wall} + \sum_{\rm modes} H_{\rm b}
  = - B \Pi + 
 \sum_{\rm modes} \left \{ B \frac{\pi^2}{2m} + \frac{Kb^2}{2} \right \}
\label{Hclassical}
\end{equation}

\section{Quantum Hamiltonian}
\label{quantumhamiltonian}

The wall Hamiltonian given by Eq.~(\ref{energyfunction})
is especially complicated because $H_{\rm wall}$ also
enters the right-hand side through $B$. In the incipient
limit, however, $H_{\rm wall}$ simplifies to the form 
Eq.~(\ref{finalHwall}) and is amenable to analysis.
Now, in Eq.~(\ref{finalHwall}), we replace
$$
R_S \rightarrow 2G H_{\rm wall}
$$
to get
$$
H_{\rm wall} = - \left ( 1-\frac{2GH_{\rm wall}}{R} \right ) \Pi
$$
and therefore
$$
H_{\rm wall} = - \left ( 1 - \frac{2G\Pi}{R} \right )^{-1} \Pi
$$
allowing us to identify
\begin{equation}
B = \frac{1}{1- 2G\Pi /R}
\label{classicalB}
\end{equation}

To quantize, we promote classical quantities to quantum operators. 
For the momentum operators we take
\footnote{
As is well-known, there are other choices for ${\hat \Pi}$,
differing by the one in Eq.~(\ref{momdef}) by functions of $R$.
These other choices will make some quantitative differences 
in the solutions given below but qualitatively the discussion
does not change.}
\begin{equation}
{\hat \Pi} = -i \frac{\partial}{\partial R} 
                            \ , \ \ \ 
{\hat \pi} = -i \frac{\partial}{\partial b}
\label{momdef}
\end{equation}
The classical Hamiltonian gets promoted to an operator and is
obtained by replacing classical variables by quantum
operators and making sure that the end result is Hermitian.
From the form of the classical Hamiltonian in
Eq.~(\ref{Hclassical}), Hermiticity can
be obtained if we choose the quantum operators corresponding
to $B\Pi$ and $B$ to be Hermitian. This is achieved by
using
\begin{equation}
{\hat B}^{-1} = 
  1- G \left ( \frac{1}{R} {\hat \Pi} + {\hat \Pi} \frac{1}{R} \right )
\end{equation}
and
\begin{eqnarray}
{\hat B} &=& 
 \left \{ 1- 
   G \left ( \frac{1}{R} {\hat \Pi} + {\hat \Pi} \frac{1}{R} \right )
 \right \}^{-1} \nonumber \\
         &=& \sum_{n=0}^{\infty}
 \left \{ G \left ( \frac{1}{R} {\hat \Pi} + 
             {\hat \Pi} \frac{1}{R} \right ) \right \}^n
\end{eqnarray}
where ${\hat \Pi}$ is defined in Eq.~(\ref{momdef}).
The Hamiltonian operator is
\begin{equation}
{\hat H} = - \frac{1}{2} ( {\hat B} {\hat \Pi} + {\hat \Pi}{\hat B}) 
           + \sum_{\rm modes} \left \{ 
 {\hat B} \frac{{\hat \pi}^2}{2m} + \frac{K}{2} b^2 \right \}
\label{quantumH}
\end{equation}
The Hermiticity of the Hamiltonian depends crucially on the 
Hermiticity of ${\hat B}$ and we discuss this further in 
Appendix~\ref{hermiticity}.

Note that ${\hat B}$ contains the inverse of the derivative 
operator and hence is really an integral operator. In 
Sec.~\ref{hatB} we will explicitly find the integral 
representation of ${\hat B}$.

With the Hamiltonian in Eq.~(\ref{quantumH}), we need to solve 
the Schrodinger equation 
\begin{equation}
{\hat H} \Psi = i \frac{\partial \Psi}{\partial t}
\label{schrodinger}
\end{equation}
where the minisuperspace wavefunction depends on $b$, $R$ and 
$t$: $\Psi = \Psi (b,R,t)$. 

We are mostly interested in the time evolution problem, where the
initial wavefunction describes a collapsing wavepacket for the 
wall and the scalar field is in its ground state. Alternately
we could study the stationary problem and seek eigenstates 
of the Hamiltonian
\begin{equation}
\Psi = e^{-iEt} \psi(b,R)
\end{equation}
This leads to the eigenvalue problem
\begin{equation}
{\hat H} \psi = E \psi
\label{HpsiEpsi}
\end{equation}
We shall discuss the stationary problem further in 
Sec.~\ref{stationary} but before doing that we find explicit
expressions for the operators ${\hat B}$ and ${\hat B}^{-1}$,
and discuss the interpretation of the wavefunction in terms
of a black hole free state.

\section{${\hat B}$ and ${\hat B}^{-1}$}
\label{hatB}

Using
\begin{equation}
[ R, \Pi ] = i
\end{equation}
we find
\begin{equation}
{\hat B}^{-1} = 1- i\frac{G}{R^2} - \frac{2G}{R}{\hat \Pi} 
\label{Bminus1}
\end{equation}

Let us define $\chi$ by
\begin{equation}
\Psi = {\hat B}^{-1} \chi
\label{Psichi}
\end{equation}
Then it is simple to find $\Psi$ in terms of $\chi$ using 
Eq.~(\ref{Bminus1})
\begin{equation}
\Psi = \left ( 1- i\frac{G}{R^2} \right ) \chi 
      + i \frac{2G}{R} \partial_R \chi 
\label{Psichiexplicit}
\end{equation}
We can also solve this differential equation to find $\chi$ in terms
of $\Psi$
\begin{equation}
\chi = {\hat B} \Psi = 
- \frac{i\sqrt{R}}{2G} e^{iR^2/4G}
\int^R dR' ~ \sqrt{R'} e^{-i{R'}^2/4G} \Psi (R')
\end{equation}
where the integral operator is indefinite. In other words, this gives 
us an explicit integral form for the ${\hat B}$ operator
\begin{equation}
{\hat B} (\cdot ) = - \frac{i\sqrt{R}}{2G} e^{iR^2/4G}
         \int^R dR' ~ \sqrt{R'} e^{-i{R'}^2/4G} (\cdot ) 
\label{Boperator}
\end{equation}
The inverse operator can similarly be written as a differential
operator
\begin{equation}
{\hat B}^{-1} (\cdot ) = 
+i \frac{2G}{\sqrt{R}} e^{+iR^2/4G} 
 \partial_R \left [ \frac{e^{-iR^2/4G}}{\sqrt{R}}  (\cdot ) \right ]
\label{B-1operator}
\end{equation}

For notational convenience, we define
\begin{equation}
\alpha (R) \equiv \frac{e^{-iR^2/4G}}{\sqrt{R}}
\end{equation}
and then
\begin{equation}
{\hat B} (\cdot ) = 
- \frac{i}{2G} \alpha^{-1} (R)
\int^R dR' ~ R' ~  \alpha (R') (\cdot )
\label{alphaBoperator}
\end{equation}
and
\begin{equation}
{\hat B}^{-1}(\cdot ) = +i\frac{2G}{R} \alpha^{-1}(R) 
               \partial_R [ \alpha (R) (\cdot ) ]
\end{equation}
It can be checked explicitly that 
${\hat B}^{-1} {\hat B} = 1 = {\hat B} {\hat B}^{-1}$.

\section{Black hole free state}
\label{bhfstate}

Since we wish to study the formation of a black hole starting with
a state that does not contain a black hole, it is important 
for us to define what we mean by a state that is
``black hole free''. At the classical level we can define a black 
hole free state by the condition $B >  0$ or $B^{-1}> 0 $. We can
lift these conditions to the quantum level by defining the operator
\begin{equation}
{\cal B} = 1 - \frac{G}{R} {\hat H}_{\rm wall}
             - G {\hat H}_{\rm wall} \frac{1}{R}
\end{equation}
A ``no black hole'' or ``black hole free'' state would only have
overlap with eigenfunctions of ${\cal B}$ whose eigenvalues
lie in the interval $(0 , \infty )$. This choice of a black
hole operator is not unique. Another possibility is
$R - 2G {\hat H}_{\rm wall}$.

In the incipient limit, we replace ${\cal B}$ by ${\hat B}$.
We will only be able to find eigenstates in this limit.
Let us now find these states by solving the eigenvalue problem
\begin{equation}
{\hat B} \xi_\beta = \beta \xi_\beta
\end{equation}
or, equivalently,
\begin{equation}
{\hat B}^{-1} \xi_\beta = \frac{1}{\beta} \xi_\beta
\end{equation}
This corresponds to the differential equation 
(see Eq.~(\ref{Psichiexplicit}))
\begin{equation}
\left ( 1- i\frac{G}{R^2} \right ) \xi_\beta
      + i \frac{2G}{R} \partial_R \xi_\beta 
= \frac{1}{\beta} \xi_\beta
\label{Beigen}
\end{equation}
The equation is solved to find the eigenfunctions
\begin{equation}
\xi_\beta = A \sqrt{R} e^{i(1 -\beta^{-1} )R^2/4G}
\label{Beigenstate}
\end{equation}
and the solution holds for a continuum of $\beta$.
Eigenstates with $\beta > 0$ are black hole free.

The overlap of the wavefunction with an eigenstate of 
${\hat B}^{-1}$ is given by the coefficient
\begin{equation}
a_\beta \equiv \langle \xi_\beta | \psi \rangle 
\label{abeta}
\end{equation}
If $a_\beta = 0$ for all $\beta \le 0$, then the state
$\psi$ is black hole free. The eigenvalue $\beta$ can
be said to quantify the ``black holeness'' of a state.

A problem with using ${\hat B}$ as the black hole operator
is that the eigenstates $\xi_\beta$ are not orthonormal
because $R$ lies in the semi-infinite interval $(0,\infty )$.
This is a familiar problem: ${\hat B}^{-1}$ contains
the operator ${\hat \Pi}$ which is like the radial momentum 
operator in quantum mechanics, and hence has no self-adjoint
extension \cite{Paz:2002}. The problem may be traced back to 
our approximation, ${\cal B} \to {\hat B}$, or equivalently in 
Eq.~(\ref{finalHwall}). 

We now consider if there are simultaneous eigenstates of the
black hole operator and the Hamiltonian. Using the expressions 
for ${\hat B}^{-1}$ (Eq.~(\ref{B-1operator})) and the Hamiltonian 
(Eq.~(\ref{quantumH})) one easily sees that 
\begin{equation}
[{\hat B}^{-1}, {\hat H}] \ne 0
\label{B-1H}
\end{equation}
This observation implies an uncertainty relation between energy
and black holeness -- if we know the energy of a state precisely
then there is uncertainty in its black holeness and if we know that 
an object is a black hole, its energy must not be precisely 
known. This ties in nicely with the usual understanding of
a black hole not as a pure state but as a thermal state.
While the value of the black holeness is uncertain for an
energy eigenstate, it may still be possible to say if a
particular energy eigenstate is black hole free because such a 
state is defined not by a single value of $\beta$ but by the 
semi-infinite interval $(0,\infty)$. In other words, the 
state of being black hole free is considerably weaker than
a state of definite black holeness, and an energy eigenstate
may be black hole free even if its black holeness is uncertain.

The commutator in Eq.~(\ref{B-1H}) can be evaluated explicitly
\[
[{\hat B}^{-1}, {\hat H}] = 
   G {\hat B} \left (  \frac{1}{R^3} - \frac{i}{R^2} {\hat \Pi} \right )
 + G \left ( \frac{1}{R^3} - \frac{i}{R^2} {\hat \Pi} \right ) {\hat B}
\]
The right-hand side is a complicated operator but can be roughly
estimated in the incipient limit using
$-{\hat B}{\hat \Pi} \sim M$,
the mass of the collapsing object. Then
\begin{equation}
[{\hat B}^{-1}, {\hat H}] \sim \frac{i}{R_S} 
\label{commB-1Hestimate}
\end{equation}
as may also be expected on dimensional grounds.

In the next section, we will discuss eigenstates of the Hamiltonian.  
These are stationary states and hence cannot resolve the 
gravitational collapse problem, which requires solving the
time-dependent problem in which the initial state is black hole
free. We will consider the gravitational collapse problem in
further detail in Sec.~\ref{gravcollapse}.

\section{Stationary states}
\label{stationary}

We now consider the Schrodinger equation, Eq.~(\ref{HpsiEpsi}), 
for a single eigenmode of the scalar field
\cite{Vachaspati:2006ki,Vachaspati:2007hr}, 
\begin{equation}
\left [ - \frac{1}{2} ( {\hat B} {\hat \Pi} + {\hat \Pi}{\hat B}) 
 + \left \{ 
 {\hat B} \frac{{\hat \pi}^2}{2m} + \frac{K}{2} b^2 \right \} \right ]
\psi = E\psi
\label{psiScheq}
\end{equation}
This is an integro-differential operator since ${\hat B}$ is
an integral operator. 

We first act on Eq.~(\ref{psiScheq}) by ${\hat B}^{-1}$ 
on the left and use
\begin{equation}
[{\hat B}^{-1}, {\hat \Pi}] = - \frac{2G}{R^3} + 
                              i \frac{2G}{R^2}{\hat \Pi}
\end{equation}
This brings Eq.~(\ref{psiScheq}) to the form
\begin{eqnarray}
\biggl [ 1-\frac{2G}{R}\biggl ( E- \frac{Kb^2}{2} \biggr ) \biggr ]
      {\hat \Pi} \psi &=& \nonumber \\
&& \hskip -3 cm
\biggl [ \frac{{\hat \pi}^2}{2m} + 
 \biggl ( \frac{Kb^2}{2} - E \biggr )  
               \biggl ( 1-i\frac{G}{R^2} \biggr ) \biggr ] \psi
 \nonumber \\
&& 
\hskip -2 cm
- \frac{2G}{R^3} \biggl [ 1-i R ~{\hat \Pi} \biggr]  {\hat B}\psi
\end{eqnarray}
To simplify the last term we use Eq.~(\ref{Bminus1}) in the form
\begin{equation}
{\hat \Pi} = \frac{R}{2G} 
\left [ - {\hat B}^{-1} + 1 - i\frac{G}{R^2} \right ]
\end{equation}
Hence
\begin{equation}
\frac{2G}{R^3} \biggl [ 1-i R ~{\hat \Pi} \biggr]  {\hat B}\psi
= \frac{i}{R}\psi - 
        \frac{i}{R} \left ( 1+\frac{iG}{R^2} \right ) {\hat B}\psi
\end{equation}
Therefore the Schrodinger equation becomes
\begin{eqnarray}
\biggl [ 1-\frac{2G}{R}\biggl ( E- \frac{Kb^2}{2} \biggr ) \biggr ]
      {\hat \Pi} \psi &=& \nonumber \\
&& \hskip -4 cm
\biggl [ \frac{{\hat \pi}^2}{2m} + 
 \biggl ( \frac{Kb^2}{2} - E \biggr )  
               \biggl ( 1-i\frac{G}{R^2} \biggr ) 
 - \frac{i}{R}  \nonumber \\
&& \hskip -1.5 cm
+ \frac{i}{R} \left ( 1 + i\frac{G}{R^2} \right ) {\hat B}
\biggr ] \psi
\end{eqnarray}

In the incipient limit ($B \to 0$) we expect the last term
to be small, say compared to the second last term ($-i/R$) 
and we drop it. Thus we obtain the following differential 
equation 
\begin{eqnarray}
\biggl [ 1-\frac{2G}{R}\biggl ( E- \frac{Kb^2}{2} \biggr ) \biggr ]
       \partial_R \psi &=& \nonumber \\
&& \hskip -4.5 cm
 - \frac{i}{2m} \partial_b ^2 \psi +
 i \biggl ( \frac{Kb^2}{2} - E \biggr )  
               \biggl ( 1-i\frac{G}{R^2} \biggr )\psi + \frac{\psi}{R} 
\label{finaldiffeq}
\end{eqnarray}

The right-hand side of Eq.~(\ref{finaldiffeq}) contains the
$G/R^2$ and $1/R$ terms arising due to the commutators of 
operators. The equation becomes intuitively obvious under
approximations that allow us to ignore these terms. The first
of these terms is small if the size of the spherical wall is
assumed to be much larger than the Planck scale. The second
term, $1/R$, is much smaller than $E$ since we have assumed 
that the mass is much larger than the Planck mass. With these
approximations, the equation reduces to
\begin{eqnarray}
\hskip -0.5 cm
\biggl [ 1-\frac{2G}{R}\biggl ( E- \frac{Kb^2}{2} \biggr ) \biggr ]
       i \partial_R \psi &=& \nonumber \\
&& \hskip - 2.5 cm
 \biggl [ E - \biggl ( - \frac{1}{2m} \partial_b ^2  +
  \frac{Kb^2}{2}  \biggr )  \biggr ] \psi 
\label{reducedfinaldiffeq}
\end{eqnarray}
Now the right-hand side is simply the total energy minus the 
simple harmonic oscillator Hamiltonian for $b$, and the left-hand
side is the usual time evolution operator if $R$ is viewed as a
time coordinate. Also, as expected, the equation has a singularity. 
However, somewhat unexpectedly, the singularity occurs at
$$
R = 2G\left ( E - \frac{Kb^2}{2} \right )
$$
and not at $2G (E-E_{\rm b})$ where the total energy in mode 
$b$, $E_{\rm b}$, includes both kinetic and potential terms. 
The reason that only the potential
term enters the location of the singularity can be traced
back to Eq.~(\ref{psiScheq}), where it is clear that 
$b$ interacts with the metric only via the kinetic term.
Multiplying that equation by ${\hat B}^{-1}$ transfers the
interaction to be only between the $Kb^2/2$ term and the wall 
momentum, ${\hat \Pi}$, since ${\hat B}^{-1}$ includes a term 
with ${\hat \Pi}$ (Eq.~(\ref{Bminus1})).

In the limit 
$$
R \to 2G\left ( E - \frac{Kb^2}{2} \right )
$$
the leading order behavior can be found by equating the 
right-hand side of Eq.~(\ref{finaldiffeq}) to zero. An
example of a non-singular function that satisfies the
differential equation to leading order in the above limit is
\begin{equation}
\psi \sim 
\exp \left [ 
 \pm i \sqrt{\frac{mR}{G} \left ( 1+i\frac{G}{R^2} \right ) 
            } ~ b \right ]
\label{leadingsoln}
\end{equation}
A more complete solution would require the wavefunction to
extend away from the singular curve in the $(R,b)$ plane.
We cannot exclude that the wavefunction could be badly
behaved in a solution which is required to satisfy certain 
boundary conditions.

\section{Gravitational Collapse}
\label{gravcollapse}

The gravitational collapse problem can now be defined. We want
to solve the time-dependent integro-differential equation
\begin{equation}
\left [ - \frac{1}{2} ( {\hat B} {\hat \Pi} + {\hat \Pi}{\hat B}) 
 + \left \{ 
 {\hat B} \frac{{\hat \pi}^2}{2m} + \frac{K}{2} b^2 \right \} \right ]
\Psi = i \frac{\partial\Psi}{\partial t}
\label{PsiScheqt}
\end{equation}
This integro-differential problem can be converted to a differential
problem exactly as for the stationary states in the previous section, 
except that $E$ must be replaced by $i\partial_t$. The differential 
equation analogous to Eq.~(\ref{finaldiffeq}) is
\begin{eqnarray}
\biggl [ 1-\frac{2G}{R}
      \biggl ( i\partial_t - \frac{Kb^2}{2} \biggr ) \biggr ]
       \partial_R \Psi &=& \nonumber \\
&& \hskip -4.75 cm
 - \frac{i}{2m} \partial_b ^2 \Psi +
 i \biggl ( \frac{Kb^2}{2} - i\partial_t \biggr )  
   \biggl ( 1-i\frac{G}{R^2} \biggr )\Psi + \frac{\Psi}{R} 
\label{diffeqt}
\end{eqnarray}
Under assumptions similar to those discussed in the previous
section, we get an equation analogous to 
Eq.~(\ref{reducedfinaldiffeq})
\begin{eqnarray}
\biggl [ 1-\frac{2G}{R}
      \biggl ( i\partial_t - \frac{Kb^2}{2} \biggr ) \biggr ]
       i \partial_R \Psi &=& \nonumber \\
&& \hskip -3.25 cm
\left [ i\partial_t - \left (
- \frac{1}{2m} \partial_b ^2 +
   \frac{Kb^2}{2}  \right )  \right ] \Psi 
\label{reduceddiffeqt}
\end{eqnarray}
Any solution of the time-dependent simple harmonic oscillator
problem will make the right-hand side vanish and will be an 
$R-$independent solution of this equation. However, since
the equation was derived assuming a large collapsing mass,
such a solution, in which all the energy resides in the
scalar radiation, cannot be taken too literally. Also,
the solution does not resolve the time-dependent gravitational 
collapse problem.  For that, we need to choose the initial 
$\Psi$ such that it represents a gravitationally collapsing 
object which is black hole free 
{\it i.e.} $a_\beta =0$ for all $\beta \le 0$ (see 
Eq.~(\ref{abeta})). With evolution, the coefficients $a_\beta$ 
will change and we are interested in finding out if the system
remains black hole free. The solution to this problem
will also allow us to track the evolution of the harmonic 
oscillator and hence the transfer of energy via radiation 
from the wall to the scalar field during quantum collapse.

Eq.~(\ref{reduceddiffeqt}) (or (\ref{diffeqt})) is a partial 
differential equation in three variables and contains mixed 
$t$ and $R$ derivatives. It can also be singular at certain points. 
These features make it hard to analyze. We hope to return to 
Eq.~(\ref{reduceddiffeqt}) in future, perhaps using numerical 
techniques. An alternative would be to consider a linear 
superposition of stationary states that match the initial 
conditions.

\section{Discussion}
\label{discussion}

We have set up a minisuperspace version of the Schrodinger
formalism for studying quantum gravitational collapse of a 
spherical domain wall in the presence of a massless scalar field 
coupled to the metric. The description automatically includes 
backreaction of the quantum radiation on the quantum dynamics 
of the domain wall. Although the passage to
minisuperspace involves a drastic truncation of the system
degrees of freedom, it still retains those that are 
relevant to describe black hole formation and evaporation. 
A clear advantage of the present approach is that the action 
for the system greatly simplifies in the interesting limit of 
an incipient black hole and raises the hope that a solution, 
even to the notorious back-reaction problem, may be within 
reach.

In the incipient limit and in the approximation that the
collapsing mass is large (see the discussion in 
Sec.~\ref{introduction}), the Hamiltonian is given by 
Eq.~(\ref{quantumH}). A striking feature of the Hamiltonian 
is that it involves the operator ${\hat B}$ which is an 
integral operator, for which we are also able to find
an explicit representation (Eq.~(\ref{alphaBoperator})).

Our analysis in the incipient limit should be sufficient to 
study the problem of quantum collapse and pre-Hawking radiation. 
However, unlike in the semiclassical approximation where the
wall radius takes on a definite value, the wavefunction is 
defined over the entire range of possible wall radii, including
very large radii. In a fuller treatment of the problem, it may 
become necessary to extend the Hamiltonian that we have found
in the incipient limit to large values of the radius. The precise 
extension is not expected to be important, as long as the incipient 
limit of the extended Hamiltonian matches the Hamiltonian found 
here.

Once we have the minisuperspace Schrodinger equation, 
we need the relevant solutions to it that represent a
solution to the gravitational collapse problem. Such
a solution would contain the fate of matter that is 
collapsing towards forming a black hole. However, the 
initial conditions have to be chosen to be ``black hole 
free'' and some measure of ``black holeness'' has to be 
defined. We have discussed some possible black hole
operators, an example of which is the operator ${\cal B}$
that coincides with ${\hat B}$ in the incipient limit. 
We find eigenstates of ${\hat B}$ and the eigenvalue 
$\beta$ may be used as a measure of black holeness. States 
with $\beta > 0$ can be said to be black hole free. However, 
the black hole operator does not commute with the Hamiltonian,
thus implying an energy-black holeness uncertainty relation.

It is interesting to note that the eigenfunction $\xi_\beta$ 
of the operator ${\hat B}$ oscillates infinitely fast
as $\beta \to 0$ which corresponds to the domain wall tending 
to the event horizon (see Eq.~(\ref{Beigenstate})). Naively, if
we start with a wavefunction that overlaps only with states with
some finite range with $\beta > 0$, to get an overlap with states
with $\beta < 0$, it would seem that we would need to go through
the infinitely oscillating state at $\beta =0 $. Quantum
mechanically, though, this is not clear. An analogy might help
clarify this situation. In the case of a non-realativistic Schrodinger
particle, our initial state may be a wavepacket having overlap
with only positive momentum eigenstates $e^{ikx}$ with $k > 0$.
Suppose at some later time, we find that there is non-zero overlap
with a negative momentum eigenstate, $e^{ipx}$ with $p < 0$.
Classically this would mean that the particle's momentum has
reversed and so, at some point in time, the particle had to pass
through zero momentum. In quantum mechanics, however, this 
is not essential since the particle need not even have a
well-defined momentum at intermediate times. So it seems that
the singular state at $\beta =0$ cannot be used to argue, at 
least straightforwardly, that there is an obstruction 
to the formation of a black hole.

While we have limited ourselves in this paper to a massless
scalar field, it is interesting to consider how the analysis
might change if the scalar field has a non-vanishing mass.
In that case, the spectrum of eigenmodes of the scalar field 
would contain bound states in addition to scattering states.
Gravitational collapse of the domain wall would then populate 
both the scattering and bound states, transferring energy from 
the wall to pre-Hawking radiation and also to an atmosphere
of self-gravitating scalar particles. Presumably, the 
bound states will result in a boson star, though the details
are not clear since boson stars as solutions of free or 
interacting scalar field theory are themselves known to have 
instabilities if their mass is 
large \cite{Ruffini:1969qy,Colpi:1986ye,Jetzer:1991jr}. 

In Sec.~\ref{stationary} we have discussed eigenstates of the
Hamiltonian and in Sec.~\ref{gravcollapse} we have set up a 
differential equation that describes the gravitational collapse 
problem. The differential equation contains mixed $t$ and $R$ 
derivatives, and it may also become singular along curves in 
the $(R,b)$ plane. These characteristics
make the equation hard enough that we have postponed
attempts at its solution for future work. In its solution,
though, might lie answers to some of the questions we have
raised in the introduction, and
we may be able to see if the unitary evolution contained in
the Schrodinger equation is self-limiting.
On the other hand, it may be that the Schrodinger equation
always yields unitary evolution, and since it is expected that 
black holes violate unitarity, it may be impossible to get to 
the black hole state as suggested by the semiclassical
calculation. 

\begin{acknowledgments}
I am grateful to Gia Dvali, Gil Paz, Al Shapere and 
Dejan Stojkovic for discussions of various aspects of 
quantum collapse and black holes. This work was supported
by the U.S. Department of Energy and NASA at Case Western 
Reserve University. 
\end{acknowledgments}

\appendix

\section{Hermiticity}
\label{hermiticity}

The Hamiltonian in the incipient limit contains the 
momentum operator but the variable $R$ lies
in the interval $(0,\infty)$. Thus ${\hat \Pi}$ resembles
the radial momentum operator which is known to have problems
with self-adjointness \cite{Paz:2002}. For us, however, 
${\hat \Pi}$ only arises in the incipient approximation and
there are issues if we blindly use this Hamiltonian over
the entire range of variables. Here we consider the 
Hermiticity of ${\hat \Pi}$ and ${\hat B}$, listing the
boundary conditions on the wavefunctions necessary to
ensure Hermiticity. 

For the momentum operator we have
\begin{equation}
\int_0^\infty dR f^\dag {\hat \Pi} g 
           = \int_0^\infty dR ({\hat \Pi} f)^\dag g
             -i [ f^\dag g ]_0^\infty \nonumber
\end{equation}
The boundary term vanishes if the functions $f(R)$ and $g(R)$ 
vanish at $R=0$ and as $R \rightarrow \infty$. So
${\hat \Pi}^\dag = {\hat \Pi}$ when acting (only) on this set 
of functions.

Next we find $({\hat B}^{-1})^\dag = {\hat B}^{-1}$ by using 
the Hermiticity of $1/R$ and ${\hat \Pi}$. This gives
\begin{eqnarray}
\int_0^\infty dR  f^\dag {\hat B}^{-1} g
 &=& \int_0^\infty dR  f^\dag 
      \left ( 1 - \frac{G}{R} {\hat \Pi} - 
               G{\hat \Pi} \frac{1}{R} \right ) g \nonumber \\
\hskip - 1cm 
 &=& \int_0^\infty dR  ({\hat B}^{-1} f )^\dag g
             +i 2G [  f^\dag g/R ]_0^\infty \nonumber
\end{eqnarray}
and so ${\hat B}^{-1}$ is also Hermitian provided $f$ and $g$
vanish sufficiently rapidly at the origin and do not grow at
infinity.

Next we find ${\hat B}^\dag$. 
\begin{eqnarray}
\int_0^\infty dR  f^\dag {\hat B} g
 &=& \int_0^\infty dR ({\hat B}^{-1} {\hat B}f)^\dag {\hat B} g 
\nonumber \\
 && 
\hskip -3 cm 
=\int_0^\infty dR ({\hat B}f)^\dag {\hat B}^{-1} {\hat B} g 
         - i 2 G[ ({\hat B}f)^\dag ({\hat B}g)/R ]_0^\infty 
       \nonumber \\
 && 
\hskip -2 cm 
= \int_0^\infty dR ({\hat B}f)^\dag  g 
         -i 2 G[ ({\hat B}f)^\dag ({\hat B}g)/R ]_0^\infty  
      \nonumber
\end{eqnarray}
Therefore ${\hat B}^\dag = {\hat B}$ on functions for which 
the boundary term vanishes. 

If acting on a space of wavefunctions such that the boundary 
terms are not zero, the operator ${\hat B}$ will not be Hermitian. 
This need not invalidate the formalism we have developed since
the explicit expression for ${\hat B}$ is only really valid
in the incipient limit, and does not remain valid at $R=0$ 
and $R\to \infty$. The Hamiltonian we have found would also
need to be extended beyond the domain of an incipient black hole.

\end{document}